\documentclass[prd,tightenlines,nofootinbib,showpacs,preprintnumbers,superscriptaddress,twocolumn]{revtex4-1} 
\usepackage{amsfonts,amsmath,amssymb,amsthm,hyperref}
\usepackage{graphicx}
\usepackage{color}

\newcommand{\be}{\begin{equation}}
\newcommand{\ee}{\end{equation}}
\newcommand{\beq}{\begin{eqnarray}}
\newcommand{\eeq}{\end{eqnarray}}

\def\I{\mathrm{i}}

\def\be{\begin{equation}}
\def\ee{\end{equation}}

\def\I{\mathrm{i}}

\newcommand\hairspace{\kern .08333em }
\newcommand{\Real}{\Re\!\hairspace\mathfrak{e}\hairspace}
\newcommand{\Imag}{\Im\mathfrak{m}\hairspace}

\begin{document}

\title{Quantum-gravity effects for excited states of inflationary perturbations}
 \author{David Brizuela}
 \email{david.brizuela@ehu.eus}
 \affiliation{Fisika Teorikoa eta Zientziaren Historia Saila, UPV/EHU,
   644 P.K., 48080 Bilbao, Spain} 
 
 \author{Claus Kiefer}
 \email{kiefer@thp.uni-koeln.de}
 \affiliation{Institut f\"ur Theoretische Physik, Universit\"{a}t zu
   K\"{o}ln, Z\"{u}lpicher Stra\ss e 77, 50937 K\"{o}ln, Germany} 
 
 \author{Manuel Kr\"amer}
 \email{manuel.kraemer@kuleuven.be}
 \affiliation{Instituut voor Theoretische Fysica, KU Leuven,
   Celestijnenlaan 200D, 3001 Leuven, Belgium} 
   
\author{Salvador Robles-P\'{e}rez}
 \email{salvador.robles@educa.madrid.org}
\affiliation{Departamento de Matem\'{a}ticas, IES Miguel Delibes, Miguel Hern\'{a}ndez 2, 28991 Torrej\'{o}n de la Calzada, Spain}
\affiliation{Estaci\'{o}n Ecol\'{o}gica de Biocosmolog\'{\i}a, Pedro de Alvarado 14, 06411 Medell\'{\i}n, Spain.}

\date{\today}

\begin{abstract}
We generalize former findings regarding quantum-gravitational
corrections arising from a canonical quantization of a perturbed FLRW
universe during inflation by considering an initial state for the
scalar and tensor perturbations that generalizes the adiabatic
vacuum state and allows us to consider the scenario that the perturbation
modes start their evolution in an excited state. Our result shows that
the quantum-gravitationally corrected power spectra get modified by
pre-factors including the excitation numbers.
\end{abstract}

\maketitle


\section{Introduction}

Looking for potentially observable effects of a candidate theory of
quantum gravity is one of the crucial tasks necessary to eventually
decide how to properly quantize gravity. Given that the standard
paradigm of the beginning of the evolution of our universe with an
inflationary phase is one of the physical scenarios with the highest
energies available to be tested by measuring the anisotropies of the
cosmic microwave background (CMB), many studies have focused on
finding quantum-gravity corrections for inflationary perturbations
that might lead to a measurable deviation of the CMB anisotropy
spectrum; see, for example,
\cite{Brizuela2018,Kiefer2012,Kamenshchik,Ashtekar2015,Castello2017,Bouhmadi2018,BM19, Martin2001}.

In the context of a canonical quantization of gravity that leads to
the Wheeler--DeWitt equation -- which can be considered as one of the
most conservative approaches to quantum general relativity
\cite{clausbook} -- one can calculate such corrections by applying a
semiclassical approximation technique to the Wheeler--DeWitt equation
of a perturbed Friedmann--Lema\^itre--Robertson--Walker (FLRW)
universe, which effectively leads to a description of the quantized
perturbations in the context of quantum field theory in curved
spacetime with corrections arising from the fact that also the
background geometry is quantized. The resulting
quantum-gravitationally corrected Schr\"odinger equation was first
derived in \cite{Kiefer1991} and applied to a simplified inflationary
model with perturbations in \cite{Kiefer2012}. In \cite{Brizuela2016a,
  Brizuela2016b} this study was extended to gauge-invariant scalar and
tensor perturbations and this is the formalism that we will base the
present study on. 

In the previous works, the initial state for the quantized
perturbations was taken to be the adiabatic vacuum state,
as it is usually done in the context of inflation; this is because of
the assumption that the initial state should correspond to the
vacuum in Minkowski space. However, given our very limited
understanding of the physics present at the onset of inflation, this
might not be the case. 

The scenario that cosmological perturbations start their evolution in
an excited state has thus already been discussed with various
motivations in the literature ranging from general considerations
\cite{Armendariz2007, Ashoorioon2014, Broy2016} to the study of non-gaussianities
\cite{Agullo2011,Ganc2011,Aravind2013}. An early discussion of initial
excited states in the Schr\"odinger picture and its semiclassical
limit can be found in \cite{LPS97}.

In this article, we intend to develop the tools needed to eventually
answer the questions whether the inclusion of a quantized background
and thus the emergence of quantum-gravity corrections can shed some
light onto the problem of choosing the initial state of inflationary
perturbations and, in particular, whether quantum-gravitational
effects might actually lead to a preference of initial excited
states. 

For this purpose, we will work with the invariants of a time-dependent
harmonic oscillator \cite{RP2017d,Lewis1969, Dantas1992, Rajeev2018}; see also
sections~8 and 20 of \cite{DR01}. The
advantage of this method is significant in that it provides a specific
criterion for considering not only the vacuum state but also the
excited states of the perturbation modes when applied to the study
presented in \cite{Brizuela2016a, Brizuela2016b}. One can then impose
an excited state as the initial state of the perturbation modes and
analyze the residual effect that it would leave in the power spectrum
of the CMB.

We have organized the article as follows. In Section II, we review the
method of the invariants of a time-dependent harmonic oscillator
leading us to the ansatz for the wave function that we then use in Section III to
compute the power spectra of the uncorrected and
quantum-gravitationally corrected perturbations for an initial excited
eigenstate, generalizing the work in \cite{Brizuela2016a,
  Brizuela2016b}. In Section IV we consider a more
general initial state given as a linear 
combination of eigenstates. We wind up the article with our
conclusions in Section V. 

\section{The invariant vacuum}

Given a harmonic oscillator with time-dependent mass and frequency, $m(t)$ and $\omega(t)$, respectively, which obeys a wave equation
\be\label{ho01}
y'' + \frac{m'}{m}\,y' +\omega^2 y = 0
\ee
with position variable $y$ and the derivative of a generic time
variable $t$ marked by a prime, it is always possible to define a
vacuum state of the quantum description that is stable along the
entire evolution of the harmonic oscillator, that is, it represents the
ground state of the harmonic oscillator along the entire evolution
irrespective of the time dependence of the Hamiltonian. This vacuum
state is called the invariant vacuum state \cite{RP2017d}.  

The Schr\"{o}dinger equation for the wave function $\psi$ that
describes the evolution of the quantum-mechanical states of the
harmonic oscillator (\ref{ho01}) is given by (setting $\hbar=1$)
\be\label{SCH01a}
H \psi = i\,\frac{\partial}{\partial t} \psi ,
\ee
with
\be\label{H}
H = \frac{1}{2m}\,\hat{p}^2 + \frac{m \omega^2}{2}\,\hat{y}^2,
\ee
where $\hat{p}$ is the momentum conjugate to $\hat{y}$. Because of the time dependence of the frequency and the mass of the harmonic oscillator (\ref{ho01}),
the diagonal representation of the Hamiltonian (\ref{H}) is not an invariant representation. This means that the vacuum state of the diagonal representation at
a given time is not the vacuum state of the diagonal representation at another given time. Instead, we look for a vacuum state that represents the ground state along
the entire evolution of the harmonic oscillator (\ref{ho01}), i.e.~a vacuum state that is invariant under the operation of time translation. The invariant
vacuum state is then given by the state that is annihilated by the
annihilation operator $\hat{b}_I(t)$ of an invariant representation of
the Hamiltonian (\ref{H}). This operator, along with its creation
counterpart $\hat{b}_I^\dag(t)$, can be defined as \cite{Lewis1969,
  Dantas1992}  
\begin{eqnarray}\label{B0}
\hat{b}_I(t) & = & \sqrt{\frac{\alpha}{2}} \left( \frac{1}{\sigma}\,\hat{y} + \frac{i}{\alpha} \left( \sigma \hat{p} - m \sigma' \hat{y} \right) \right) , \\ \label{B0d}
\hat{b}_I^\dag(t) & = & \sqrt{\frac{\alpha}{2}} \left(
                        \frac{1}{\sigma}\,\hat{y} -  \frac{i}{\alpha}  \left( \sigma
                        \hat{p} - m \sigma' \hat{y} \right) \right) , 
\end{eqnarray}
where $\alpha$ is a constant (see (\ref{sigmaequation}) below) and $\sigma \equiv \sigma(t)$ is a function to be determined later on.
The operators obey the commutation relation $[\hat b_I , \hat b_I^\dag] = 1$ and fulfill
\beq
\hat b_I^\dag |N\rangle &=& \sqrt{N+1} |N+1\rangle , \\ 
\hat b_I |N\rangle &=& \sqrt{N} |N-1\rangle ,
\eeq
with $|N\rangle$ being the number state of the invariant representation, which is generated as usual from the successive application of the creation operator $\hat b_I^\dag$ upon the vacuum state of the invariant representation,
\be
|N\rangle = \frac{\bigl(\hat b_I^\dag\bigr)^N}{\sqrt{N!}} |0\rangle .
\ee
The main property of the invariant representation is precisely that it defines an invariant, namely the operator $\hat N_I \equiv \hat b_I^\dag \hat b_I$, which satisfies  
\be\label{INV01}
\frac{d}{d t} \!\left(\hat{b}_I^\dag \hat{b}_I\right) = \frac{\partial \bigl(\hat{b}_I^\dag \hat{b}_I\bigr)}{\partial t}  + i \left[ H, \hat{b}_I^\dag \hat{b}_I \right] =0,
\ee
such that 
\be \label{numop}
\hat{N}_I(t)  |N \rangle_I \equiv \hat{b}_I^\dag \hat{b}_I |N \rangle_I = N  |N\rangle_I ,
\ee 
with $N$ being a constant integer number. It means that once the harmonic oscillator is in a number state of the invariant representation it remains in the same state along the entire evolution. In particular, once it is in the invariant vacuum state it remains in the vacuum state for all time.

From (\ref{INV01}) it is clear that the Hamiltonian (\ref{H}) cannot be diagonal in the invariant representation because in this case the commutator in (\ref{INV01}) would be zero and the total derivative would not vanish.
Furthermore, the function $\sigma$ cannot be a constant because this would imply that the partial derivative vanishes, but not the commutator at any time. The exact form of the Hamiltonian can be easily obtained by taking $\hat y$ and $\hat p$ from (\ref{B0}--\ref{B0d})
and inserting them into (\ref{H}). It yields
\be\label{H2}
H= \beta_- \hat b_I^2 + \beta_+ (\hat b_I^\dag)^2 + \beta_0 \left( \hat b_I^\dag \hat b_I + \frac{1}{2} \right),
\ee 
where $\beta_\pm$ and $\beta_0$ are time-dependent functions, which we do not write out here explicitly for reasons of compactness.
The Hamiltonian (\ref{H2}) is known to generate squeezed states. From that point of view, the number states of the invariant representation (see (\ref{numop})) can be seen as the squeezed number states whose eigenvalues exactly remain constant along the evolution of the harmonic oscillator (\ref{ho01}).
The precise form of the function $\sigma(t)$ is the one that ensures that (\ref{INV01}) is satisfied. This turns out to be the modulus of the solution to the classical evolution equation \eqref{ho01}, which can be written using a real phase $\tau$ as 
$$y(t)\equiv\sigma(t) e^{i\tau(t)}.$$
By replacing this polar decomposition into \eqref{ho01}, it can be easily shown that $\sigma$ 
satisfies the auxiliary equation \cite{Lewis1969}
\be\label{sigmaequation}
\sigma'' + \frac{m'}{m} \sigma' + \omega^2   \sigma = \frac{\alpha^2}{m^2 \sigma^3},
\ee
with an arbitrary real constant $\alpha$ that is related to the
normalization of the modes, 
while the phase is written as a quadrature,
\be \label{TAU01a}
\tau(t) = \int^t \frac{\alpha}{m(\tilde t) \sigma^2(\tilde t)} \,d\tilde t.
\ee
In order to fix the rest of the integration constants,
it is usual practice to look for an asymptotic limit where the mass
and frequency of the 
harmonic oscillator are constants, given by $m_0$ and $\omega_0$, respectively.
In this limit, the invariant representation (\ref{B0}--\ref{B0d}) has to become the customary representation
of the harmonic oscillator with constant mass and frequency. This is accomplished
by choosing the modes to have a constant amplitude,
\be\label{BC011}
\sigma \rightarrow\sqrt{\frac{\alpha}{m_0 \omega_0}}  \ , \quad \sigma' \rightarrow 0,
\ee 
in that limit. Note that the value of the amplitude has been chosen such that $\sigma''$
is vanishing, as can be seen from its equation of motion \eqref{sigmaequation}. This makes the representation
unchanged as long as the mass and frequency keep their constant values.
These conditions translate to the following asymptotic form for the classical solution,
\be
y\rightarrow \sqrt{\frac{\alpha}{m_0 \omega_0}}\,e^{i\omega_0 t}, \quad |y|'\rightarrow 0.
\ee
At this point, the normalization $\alpha$ is the only constant to be fixed and, for that, one usually
considers the Wronskian defined using the complex conjugate $\overline{y}$ as
\begin{equation}\label{wronskian}
W[\overline{y}, y]= \overline{y} y'- \overline{y}'y =\frac{2 i \alpha}{m},
\end{equation}
which is a constant of motion.
We note that the constant $\alpha$ has the dimension of an action. Since
we use here units in which $\hbar=1$ (in addition to $c=1$), $\alpha$
is dimensionless and we 
can set for convenience $\alpha=1$. (Alternatively, one could re-scale
$\alpha\to\alpha\hbar$.)

As usual, the solutions of the Schr\"odinger equation \eqref{SCH01a} correspond to the eigenstates of the number operator \eqref{numop}; see, for instance, Refs.~\cite{Leach1983, Kanasugui1995}. They are given by 
the following wave function \cite{Leach1983}: 
\be \label{NS01a}
\psi_{N}^{\rm eig}(y,t) \equiv \langle y | N \rangle =  \frac{1}{\sqrt{\sigma(t)}} \exp\!\left( \frac{i\, m}{2}\frac{\sigma'(t)}{\sigma(t)} \,y^2 \right) {\varphi}_N(y,t),
\ee
where $\varphi_N$ is the customary wave function of the $N$th eigenstate of the
harmonic oscillator with unit mass and frequency, that is,
\be
\varphi_N(y,t) = \frac{e^{-i (N+\frac{1}{2}) \tau(t)}}{\sqrt{2^N N!}\,\pi^\frac{1}{4}}\,e^{-\frac{y^2}{2\sigma^2(t)}}  \,{\rm H}_N\!\left(\frac{y}{\sigma(t)}\right) ,
\ee
with ${\rm H}_N$ being the Hermite polynomial of degree $N$, and $\tau(t)$ is the
phase of the solution to the classical equation defined in \eqref{TAU01a}.

In particular, the wave function of the vacuum state ($N=0$) for the harmonic oscillator with
time-dependent frequency and mass reads
\be\label{VWF01}
\psi_{0}^{\rm eig}(y,t) = \frac{e^{-\frac{i}{2}\tau(t)}}{\sqrt{\sigma(t)}\,\pi^\frac{1}{4}} \exp\!\left[-\,\frac{1}{2} \left(\frac{1}{\sigma^2(t)} - i\, m\,\frac{\sigma'(t)}{\sigma(t)} \right) y^2 \right] ,
\ee
which is a Gaussian wave function with variance $\sigma^2(t)$. (The
above constant $\alpha$, which was set equal to one, can be recovered
by the substitution $\sigma \to\sigma/\sqrt{\alpha}$.)

\section{Power spectra for an initial excited eigenstate}

In this section, we will use the method explained above in order to obtain the power spectra
corresponding to both scalar and tensorial perturbations, which can be related to the anisotropies
seen in the cosmic microwave background.
We will do so by first following the
standard approximation given by quantum field theory on classical background spacetimes
and then by including quantum-gravity effects coming from the quantization of that background
spacetime. In both cases, the equation to be solved will correspond to a Schr\"odinger equation
of a harmonic oscillator with time-dependent frequency and unit mass. 

{For the scalar and tensor perturbations of the FLRW spacetime that we are considering, we use the perturbation variables $v_k$ that
correspond to the Mukhanov--Sasaki variables \cite{Mukhanov1988} in the scalar sector and to the two different independent components of the symmetric tensor that
describes the gravitational waves in the tensorial sector. The exact
definitions can be found, for example, in \cite{Brizuela2016a,Brizuela2016b}. 
The power spectrum ${\cal P}_v$ of the variable $v_k$ can then be defined using the wave function \eqref{NS01a} for a number eigenstate $N_k$, where $N_k$ is an integer, with $v_k$ taking the role of the position variable and the conformal time $\eta$ as the time variable, as follows (see e.g.~\cite{Martin2012}):
\begin{eqnarray}
& & \frac{2\pi^2}{k^3}\, {\cal P}_v(k)\,\delta(k-p)=\langle \hat v_k
                                                   \overline{\hat{v}}_p\rangle\\ 
& & \ =\int \prod_qdv_q\,\overline{\psi}_{N_q}^{\rm
    eig}(v_q,\eta)\,\hat{v}_k\overline{\hat{v}}_p\,\psi_{N_q}^{\rm
    eig}(v_q,\eta). \nonumber 
\end{eqnarray} 
Evaluating the integral above
then leads to
\begin{equation}\label{psv}
 {\cal P}_v(k) = \frac{k^3}{2\pi^2}\frac{(2 N_k+1)}{2}\,\sigma^2.
\end{equation}
The quantity $\sigma(k,\eta)$ is the modulus of the mode variable $y(k,\eta)$ that obeys an equation analogous to \eqref{ho01}.

The real observable power spectrum of the scalar sector corresponds to
the spectrum of the
comoving curvature perturbation $\zeta$, which is related to the
Mukhanov--Sasaki variable as follows,
\be
\label{eq:linkzetav}
\zeta_{k} =\sqrt{\frac{4\pi G}{\epsilon}}\,\frac{v_{k}}{a}.
\ee
The power spectrum of the scalar sector is thus given by
\be\label{scalarps}
{}^{\rm S}{\cal P}_{N_k}(k):= \frac{4\pi G}{a^2\,\epsilon}\,\frac{k^3}{2\pi^2}\,\frac{2N_k+1}{2}\,\sigma^2.
\ee
Concerning the tensorial sector, the power spectrum is given by
\be
\label{tensorialps} 
{}^{\rm T}{\cal P}_{N_k}(k) :=\,\frac{64\pi G}{a^2}\,\frac{k^3}{2\pi^2}\,\frac{2N_k+1}{2}\,\sigma^2.
\ee
In both cases, $\sigma(k,\eta)$ should be obtained by solving its equation of motion \eqref{sigmaequation}
with its corresponding frequency.

\subsection{Uncorrected perturbation modes}

Following \cite{Brizuela2016a, Brizuela2016b}, the wave function of the uncorrected perturbation modes, $\psi_{k}^{(0)}$, is the solution of the Schr\"{o}dinger equation
\be\label{SCh01}
\mathcal{H}_k \psi_{k}^{(0)} = \I\,\frac{\partial }{\partial \eta} \psi_{k}^{(0)} ,
\ee
with
\be\label{H02}
\mathcal{H}_k = -\,\frac{1}{2}\,\frac{\partial^2}{\partial v_{k}^2}+\frac{1}{2}\,\omega^2(k,\eta)\,v_{k}^2 .
\ee
It corresponds to the Hamiltonian of a harmonic oscillator with constant mass, $m= 1$, and frequency given by
\be
\label{defomega}
\omega^2_\text{S}(k,\eta):=k^2-\frac{z^{\prime\prime}}{z}\,,\quad
\omega^2_\text{T}(k,\eta):=k^2-\frac{a^{\prime\prime}}{a}\,,
\ee
for the scalar modes, $\omega_\text{S}$, and for the tensorial modes, $\omega_\text{T}$.
The prime now stands for the derivative with respect to the conformal
time $\eta$, while $a=a(\eta)$ is the scale factor 
and $z:=a\,\phi^\prime/\mathcal{H}$, with $\phi$ being the inflaton field and $\mathcal{H} := a'/a$ (see Refs.~\cite{Brizuela2016a, Brizuela2016b} for the details).

For simplicity, in this paper we will just
work in the de Sitter limit with a constant Hubble factor $H$. In that case both scalar and tensorial frequencies
coincide and take the following form,\footnote{We will use the
  subindex $0$ to refer to objects related to the uncorrected wave
  function $\psi^{(0)}_k$, 
like $\omega_0$, $\sigma_0$ and $y_0$, whereas the subindex $1$ will
stand for objects associated with 
the corrected wave function $\Psi^{(1)}_k$ that will be introduced in
the next subsection.} 
\be \label{omega0}
\omega_{0}^2(k,\eta)=k^2-\frac{2}{\eta^2}.
\ee
In order to obtain the power spectrum, one assumes that the wave function
is in a given eigenstate $N_k$ and sets $\psi_{k}^{(0)}(v_k,\eta)\equiv\psi_{N_k}^{\rm eig}(v_k,\eta)$ using \eqref{NS01a}. Then one just needs to solve the equation
for the mode variable $y_{0}(k,\eta)$ and insert its modulus $\sigma_{0}(k,\eta)$ into \eqref{scalarps} and \eqref{tensorialps}. 
The equation for the modes $y_{0}$ takes the form
\be \label{yuncorr}
y_{0}'' + \omega_{0}^2\,y_{0} = 0 ,
\ee
and can be solved straightforwardly by
\be\label{genericsoly0}
y_{0}=c_1 \left(\frac{\sin\xi}{\xi}-\cos\xi \right)+c_2 \left(\frac{\cos\xi}{\xi}+\sin\xi \right),
\ee
where, for convenience, we have defined the new dimensionless time
variable $\xi:=-k\eta$ and have introduced two integration
constants $c_1$ and $c_2$. 

As stated in the previous section, in order to fix the integration constants,
one should find a limit where the frequency tends to a certain constant. In inflationary
dynamics, this happens at the beginning of inflation $\eta\rightarrow-\infty$ (or
equivalently $\xi\rightarrow\infty$). In that limit, we have $\omega_{0}\rightarrow k$, so one
gets the modes corresponding to Minkowski spacetime. Therefore, we will apply
the conditions
\be
y_{0}\rightarrow \frac{e^{-i\xi}}{\sqrt{k}},\qquad|y_{0}|'\rightarrow 0.
\ee
With these conditions, and the normalization of the Wronskian \eqref{wronskian}, one gets
the solution
\be\label{soly0}
y_{0}=\frac{\xi-i}{\sqrt{k}\,\xi}\,e^{-i\xi},
\ee
whose modulus is given by
\be\label{sigma0}
\sigma_{0}=\left(\frac{\xi^2+1}{k\,\xi^2}\right)^{1/2}.
\ee
Inserting this result into the power spectra \eqref{scalarps} and \eqref{tensorialps}
and taking into
account that the scale factor behaves as $a=-1/(H\eta)=k/(H\xi)$, one gets
the usual scale-invariant form for the power spectra for super-Hubble scales
($\xi\rightarrow 0$),
\begin{eqnarray}
{}^{\rm S}{\cal P}^{(0)}_{N_k}(k)&=&(2N_k+1)\,\frac{G H^2}{\pi\epsilon}, \\
{}^{\rm T}{\cal P}^{(0)}_{N_k}(k)&=&(2N_k+1)\,\frac{16 G H^2}{\pi}.
\end{eqnarray}
Note that in these objects we have included the superindex $(0)$ to explicitly mark that they correspond
to the uncorrected case. In the limit $N_k\to 0$ we recover the
results of Ref.~\cite{Brizuela2016a} for the power spectra corresponding to the ground state.

\subsection{Corrected perturbation modes}

The consideration of a non-vacuum state for the initial state of the perturbations of the scalar field or the spacetime would have two consequences in the next-order corrected wave functions $\psi_{k}^{(1)}$ computed in \cite{Brizuela2016a, Brizuela2016b}. One is the modification that the choice of a non-vacuum state introduces in the computation of the corrected frequency $\widetilde \omega$ (see below). The other is that one can also apply the method of the invariants of the harmonic oscillator to obtain the excited states of the corrected wave functions, {$\psi_{k}^{(1)}(v_k,\eta) \equiv \psi_{\widetilde{N}_k}^{\rm eig}(v_k,\eta)$ using \eqref{NS01a}, and check their influence in the generated power spectrum.

The Schr\"{o}dinger equation for the perturbation modes corrected by the quantum-gravitational effects of the Wheeler--DeWitt equation is given by \cite{Brizuela2016a, Brizuela2016b}
\begin{eqnarray} \label{corrSchreq} 
&&i\,\frac{\partial}{\partial \eta}\,\psi_{k}^{(1)} =
\mathcal{H}_{k}\psi_{k}^{(1)} \\
&-&\frac{\psi_{k}^{(1)}}{2\,m_{\rm P}^2}\,
\Real\!\left\{\frac{1}{\psi_{k}^{(0)}}\Biggl[\frac{\bigl(\mathcal{H}_{k}\bigr)^2}{V}\,\psi_{k}^{(0)}
+ i\,\frac{\partial}{\partial \eta}\!\left(\frac{\mathcal{H}_{k}}{V}\right)\psi_{k}^{(0)}\Biggr]\right\}\,. \nonumber
\end{eqnarray}
Here we use $m_{\rm P}$ to denote a rescaled Planck mass $m_{\rm P} :=
3/(4\pi G)$ (recall $\hbar=1=c$)
and $V$ is a minisuperspace potential that in de Sitter space takes the form
\be \label{V0}
V(\eta) = \frac{1}{H^2\eta^4}.
\ee
In order to avoid unitarity violations, and following the considerations presented in
\cite{Brizuela2016a, Brizuela2016b}, we have taken the real part of the term between the curly brackets in (\ref{corrSchreq}),
see Refs.~\cite{Kiefer2018, Chataignier2018}. In order to compute that term, we will use the ansatz
$\psi_{k}^{(0)}(v_k,\eta)=\psi_{N_k}^{\rm eig}(v_k,\eta)$ and perform a power expansion in $v_k$ of the result,
dropping terms of order $v_k^4$ and higher.
Note that this power expansion does not lead to any odd power because of the
parity of the eigenfunctions $\psi_{N_k}^{\rm eig}$. In this way, the above equation is rewritten
as follows,
\begin{eqnarray} \label{corrSchreq2} 
&&i \frac{\partial}{\partial \eta}\,\psi_{k}^{(1)} =
\mathcal{H}_{k}\psi_{k}^{(1)}
+\left(f_0+v_k^2 f_2\right)\psi_{k}^{(1)},
\end{eqnarray}
where $f_0$ and $f_2$ are real functions of time that depend on
$\sigma_{0}$, $\omega_{0}$, $V$ and the number of the state $N_k$, but not on $v_k$. In fact, these functions differ
whether $N_k$ is an even or odd number. Explicitly, they are given by:
\begin{eqnarray*}
f_0&=& \frac{1}{8 m_{\rm P}^2 V}\left[\frac{3 \sigma_0 '^2}{\sigma_0^2}+2 \omega_0^2-\frac{3+4 N_k (N_k+1)}{\sigma_0^4}+\frac{2 \sigma_0' V'}{\sigma_0 V}\right],\\
f_2&=& \frac{2 N_k+1}{4m_{\rm P}^2 V\sigma_0^6}\left[3 \left(1-3 \sigma_0^2 \sigma_0'^2-\sigma_0^4 \omega_0^2\right)
-\frac{2\sigma_0^3 \sigma_0' V'}{V}\right],
\end{eqnarray*}
for even $N_k$, and
\begin{eqnarray*}
f_0 &=&\frac{1}{8 m_{\rm P}^2 V} \left[\frac{15 \sigma_0'^2}{\sigma_0^2}+6 \omega_0^2
-\frac{7+4 N_k(N_k+1)}{\sigma_0^4}+\frac{6 \sigma_0 ' V'}{\sigma_0 V}\right], \\
f_2 &=&\frac{2N_k+1}{12 m_{\rm P}^2 V\sigma_0^6}\left[
5 \left(1-3 \sigma_0 ^2 \sigma_0'^2-\sigma_0^4 \omega_0^2\right)
-\frac{2 \sigma_0^3 \sigma_0' V'}{V}
\right],
\end{eqnarray*}
for odd $N_k$.

At this point, it is useful to redefine the wave function as 
\be
\Psi_{k}^{(1)}:=e^{-i\int^{\eta} f_0(\tilde\eta)d\tilde\eta}\,\psi_{k}^{(1)},
\ee
which just differs from the original wave function by a phase
and thus does not change the expression for the power spectrum. This new wave function
obeys the Schr\"odinger equation of a time-dependent harmonic oscillator,
\begin{equation}\label{eqpsi1tilde}
i\,\frac{\partial\Psi_{k}^{(1)}}{\partial\eta}=
- \frac{1}{2}\,\frac{\partial^2 \Psi_{k}^{(1)}}{\partial v_k^2}
+\frac{1}{2}\,{\omega}^2_{1}\,v_k^2\,\Psi_{k}^{(1)},
\end{equation}
with the modified frequency $\omega_1(k,\eta)$ given by
\be
\omega^2_{1}:=\omega_{0}^2+2 f_2.
\ee
By inserting} the form of $\omega_{0}$ \eqref{omega0} and $V$ \eqref{V0} for de Sitter spacetime
and the solution for $\sigma_{0}$ \eqref{sigma0} in the expressions above, one gets
the following form of the modified frequency:
\begin{equation}
\omega^2_{1}=k^2-\frac{2 k^2}{\xi ^2}+
            \frac{(2 N_k+1)}{k}\,\frac{H^2}{m_{\rm P}^2}\,\widetilde\omega_{1}^2,
\end{equation}
where $\widetilde\omega_{1}$ has a different expression depending on whether
$N_k$ is an even or odd number:
\begin{equation}
\widetilde\omega_1^2=\left\{
        \begin{array}{ll}
            \frac{1}{2}\frac{\left(\xi ^2-11\right) \xi ^4}{\left(\xi
   ^2+1\right)^3} & \quad \text{for } N_k \text{ even}, \\
\frac{1}{6}\frac{\left(7 \xi ^2-13\right) \xi ^4}{\left(\xi
   ^2+1\right)^3} & \quad \text{for } N_k \text{ odd}.
        \end{array}
    \right.
\end{equation}
It is interesting to note that the form of the corrected frequency
is quite similar for both even and odd cases. They both have a global
factor $(2N_k+1)/k$ multiplying the correction terms of
the frequencies, such that the dependence on $N_k$ and $k$ of the correction
to the power spectra will be the same for both even and odd $N_k$.
In addition, the corrective term is given by the ratio between
two polynomials in the dimensionless time $\xi$. These polynomials
are of the same order in both cases, but just with slightly different
coefficients. This implies that the asymptotic limits of the
frequency will be quite similar in both cases.

The next step will be to solve equation \eqref{eqpsi1tilde}. This can
be done following the steps of the previous section by
assuming that $\Psi_{k}^{(1)}$ takes a form
\be \label{exansatz}
\Psi_{k}^{(1)}(v_k,\eta) \equiv \psi_{M_k}^{\rm eig}(v_k,\eta)
\ee
based on \eqref{NS01a}. Note that the excitation number
$M_k$ used here does not necessarily need to be the same
as $N_k$, the one used when inserting $\psi_{k}^{(0)}$ into
\eqref{corrSchreq}. Nevertheless, in this framework, both
$\psi_k^{(0)}$ and $\psi_k^{(1)}$ are approximations up to different orders in $m_{\rm P}$
to the exact wave function that obeys the full Wheeler--DeWitt equation \cite{Brizuela2016a,Brizuela2016b}.
In fact, up to a certain order in $m_{\rm P}$,
$\psi_{k}^{(0)}=\psi_{k}^{(1)}$ holds. Therefore, it
seems reasonable to assume that they are in the same number eigenstates
$N_k$; for this reason, we set in the following
$M_k=N_k$.

In this way,
one just needs to solve the equation for the corresponding mode variable $y_1(k,\eta)$,
\begin{equation}\label{correctedequation}
 y_{1}''+\omega_{1}^2\,y_{1}=0,
\end{equation}
with the following conditions at the beginning of inflation $\eta\rightarrow-\infty$,
\begin{equation}
 y_{1}\rightarrow\frac{e^{i\,\omega_{1}(\infty)\,\eta}}{\sqrt{\omega_{1}(\infty)}},\qquad |y_{1}|'\rightarrow 0,
\end{equation}
where $\omega_{1}(\infty)$ is the value of $\omega_{1}$ in that limit,
and the normalization $W[\overline{y}_{1},y_{1}]=2 i$. The modulus $\sigma_{1}:=|y_{1}|$
should then simply be plugged into the form of the power spectra \eqref{scalarps}--\eqref{tensorialps}.

Equation \eqref{correctedequation} turns out to be quite difficult to solve
analytically. Therefore, we will perform the following change of variable,
\begin{equation}
 y_{1} \equiv y_{0}+\frac{2N_k+1}{k^{7/2}}\,\frac{H^2}{m_{\rm P}^2}\,\widetilde{y}_{1},
\end{equation}
where $y_{0}$ is the solution to the uncorrected equation \eqref{yuncorr} and is explicitly
given by \eqref{soly0}. Dropping terms of the order $(H/m_{\rm P})^4$, the equation
for $\widetilde y_{1}$, written in terms of $\xi$, is given by
\begin{equation}
 \frac{d^2\widetilde y_{1}}{d\xi^2}+\frac{\omega_{0}^2}{k^2}\,\widetilde y_{1}= -\sqrt{k}\,y_{0}\, {\widetilde\omega_1^2},
\end{equation}
which is the equation of the uncorrected harmonic oscillator with an
inhomogeneous term. The general solution is then given by the solution
of the homogeneous part \eqref{genericsoly0} plus some particular solution of
the inhomogeneous equation, which can be systematically obtained.
We will do so in the following for both even and odd cases.
In addition, all the $k$-dependence of this equation is encoded in $\xi$.
Note that the explicit $k$ that are written in that expression are
absorbed by the dependence on $k$ of $\omega_{0}^2$ and $y_{0}$.
The modulus of the mode function $y_{1}$ is then given by
\begin{equation}
\sigma^2_{1}=\sigma_{0}^2+\frac{2 N_k+1}{k^{4}}\,\frac{H^2}{m_{\rm P}^2}\,\widetilde\sigma^2_1,
\end{equation}
where we have defined
\begin{equation}
\widetilde\sigma^2_1:= 2\sqrt{k}\,\Real\!\left(y_{0}\,{\overline{\widetilde{y}}}_1\right).
\end{equation}
This last $\sqrt{k}$ has been inserted to compensate the $k$-dependence of $y_{0}$
and make, in this way, $\widetilde\sigma^2_1$ independent of $k$.
The initial conditions for $\widetilde y_1$ can be obtained by expanding
in a power series of $(H/m_{\rm P})^2$ the above conditions for $y_{1}$, and they
are given in a compact way in terms of this last object as follows,
\begin{equation}
\widetilde\sigma^2_1\rightarrow-\frac{\widetilde\omega_1(\infty)^2}{2},\qquad (\widetilde\sigma_1^2)'\rightarrow 0,
\end{equation}
$\widetilde\omega_1(\infty)$ being the value of the function $\widetilde\omega_1$ at $\xi\rightarrow\infty$.
Note that these conditions are also $k$-independent, since $\widetilde\omega_1$ is $k$-independent,
so all $k$-dependence is explicitly displayed. Finally, the normalization of the
Wronskian \eqref{wronskian} implies that
\begin{equation}
 \Imag(\overline{y}_{0}\,\widetilde y_1'+y_{0}'\,\overline{\widetilde y}_1)=0.
\end{equation}
As a side remark, we note that these conditions fix completely $\widetilde\sigma^2_1$,
but not $\widetilde y_1$, which still has some freedom. In particular,
given a $\widetilde y_1$ that obeys the conditions above,
one could add to it a term of the form $i\gamma y_{0}$ with any real constant $\gamma$,
and this new $\widetilde y_1$ would still obey the above conditions.
This does not affect the form of $\widetilde\sigma_1^2$, since its contribution
to $\sigma_{1}$ is of order $(H/m_{\rm P})^4$, which we are neglecting.
Therefore, this freedom is just an artifact of the level of approximation
we are using.

From these results, we can already display the functional form of the
corrected power spectra for both scalar and tensor perturbations,
\begin{equation} \label{corrPraw}
{\cal P}^{(1)}_{N_k}(k)={\cal P}^{(0)}_{{N}_k}(k)\left(1+\frac{2 N_k+1}{k^3}\,\frac{H^2}{m_{\rm P}^2}\,\beta_{N_k}\right),
\end{equation}
where $\beta_{N_k}$ is a number given by
\begin{equation}
 \beta_{N_k}=\lim_{\xi\rightarrow\infty} \xi^2\widetilde\sigma_1^2.
\end{equation}
In fact, the only dependence of this number will be on $N_k$, whether it is even or odd.
The calculation to obtain its explicit form for both cases is
presented in the Appendix,
and the result reads 
\begin{equation}
 \beta_{N_k}\approx\left\{
        \begin{array}{ll}
        0.9876 & \quad \text{for } N_k \text{ even}, \\
		0.104 & \quad \text{for } N_k \text{ odd}.
        \end{array}
    \right.
\end{equation}

As can be seen from \eqref{corrPraw}, we recover again the $1/k^3$ dependence of the corrected power spectra as in \cite{Kiefer2012,Brizuela2016a,Brizuela2016b} and for even $N_k$, the number $\beta_{N_k}$ corresponds to the value computed in \cite{Brizuela2016a}.
In addition, the relative correction term appears multiplied by a factor $(2N_k+1)$,
in addition to another $(2{N}_k+1)$ global factor in front of the power spectrum that arises from the ansatz \eqref{exansatz} for the corrected wave function $\Psi_{k}^{(1)}$.

Our final result for the quantum-gravitationally corrected scalar and tensor power spectra in de Sitter space using excited initial states thus reads:
\begin{eqnarray*}
{}^{\rm S}{\cal P}^{(1)}_{N_k}(k)&=&\,\frac{G H^2}{\pi\epsilon}\,(2{N}_k+1)\left(1+\frac{2 N_k+1}{k^3}\,\frac{H^2}{m_{\rm P}^2}\,\beta_{N_k}\right), \\
{}^{\rm T}{\cal P}^{(1)}_{N_k}(k)&=&\frac{16 G
                                           H^2}{\pi}\,(2{N}_k+1)\left(1+\frac{2
                                           N_k+1}{k^3}\,\frac{H^2}{m_{\rm
                                           P}^2}\,\beta_{N_k}\right). 
\end{eqnarray*}
Again, the results of \cite{Brizuela2016a} for the case of an adiabatic ground state are recovered in the limit $N_k\to 0$. 

We have not considered here the issue of the quantum-to-classical
transition for the primordial modes. This is of importance, because
these modes are of a fundamentally quantum nature, whereas the observed
CMB anisotropies can be described by classical stochastic
quantities. The mechanism for the quantum-to-classical transition is
environmental decoherence.
For the uncorrected wave functions, decoherence has been
addressed for the case of an initial ground state as well as initial
excited states \cite{KP98}. Using the explicit expressions for the
wave functions presented in
\cite{LPS97}, it was found that decoherence for number eigenstates is
stronger by a factor $2N_k+1$; see Eq.~(84) in \cite{KP98}.
The quantum-to-classical transition thus proceeds faster than for the
case of the ground state.

\section{Power spectra for a superposition of states}

In the previous section, we have computed the power spectra by considering an initial eigenstate
for the uncorrected $\psi^{(0)}_k$ and corrected wave functions $\Psi^{(1)}_k$. In order to generalize
this result, in this section we will instead assume a linear combination of those eigenstates,
that is, $\psi=\sum_{N_k}C_{N_k}\psi_{N_k}^{\rm eig}$.
In particular, with this choice of wave function, the form of the power spectrum \eqref{psv} will change
and take the following form:
\begin{eqnarray}
\mathcal{P}_v&=&\frac{k^3}{2\pi^2}\sum_{N_k,M_k} C_{N_k} \overline{C}_{M_k} \langle\psi_{N_k}^{\rm eig}|v_k \overline{v}_k|\psi_{M_k}^{\rm eig}\rangle \nonumber\\
&=&\frac{k^3\sigma^2}{4\pi^2} \sum_{N_k} \Big[(2 N_k +1)|C_{N_k}|^2\nonumber\\&+& 2 \Real(C_{N_k} \overline{C}_{{N_k}+2}) \sqrt{({N_k}+2)(N_k+1)}\Big].
\end{eqnarray}
Note that the change only affects the global factor $(2 N_k+1)$ in \eqref{psv}, which is now replaced
by the summation in the equation above, but the power spectrum is still proportional to $\sigma^{2}$.
Concerning the power spectra for scalar and tensorial modes \eqref{scalarps}--\eqref{tensorialps},
they will also change in the same way, by just replacing the factor $(2N_k+1)$ by the sum above.

Finally, since in the uncorrected case the computation of $\sigma_0$ does not depend on the number state $N_k$,
one obtains the same result \eqref{sigma0}. Nonetheless, regarding the computation
of the corrected $\sigma_1$, the corrected Schr\"odinger equation is
nonlinear in $\psi^{(0)}_k$. 
Therefore, if one assumes a linear combination of eigenstates for $\psi^{(0)}_k=\sum_{N_k}C_{N_k}\psi_{N_k}^{\rm eig}$,
different modes will couple and the corrected frequency $\omega_1$ will have an intricate dependency on
the different numbers $N_k$ and coefficients $C_{N_k}$. Although lengthy, the computation is straightforward
for a finite number of modes, but obtaining the result for the generic (in principle infinite) linear combination does not
seem feasible.

As an example, we include here the form of the corrected frequency for the particular case
$\psi^{(0)}_k=C_{1}\psi_{1}^{\rm eig}+C_{3}\psi_{3}^{\rm eig}$ for the de Sitter spacetime:
\begin{eqnarray*}
  &&\omega^2_{1}=k^2-\frac{2 k^2}{\xi ^2}\\
  & & +\frac{H^2 \xi ^4}{24 k\, m_{\rm P}^2 \left(\xi
   ^2+1\right)^3 \left(2 C_{1}^2-2 \sqrt{6} C_{1} C_{3}+3
   C_{3}^2\right)^2}\times\\
 &&  \left[12 C_{1}^4 \left(7 \xi ^2-13\right)-16 \sqrt{6} C_{1}^3
    C_{3} \left(5 \xi ^6+14 \xi ^2-26\right)\right.
  \\ && +60C_{1}^2 C_{3}^2\left(8 \xi 
   ^6+21 \xi ^2-39\right)\\ & & \left.-24 \sqrt{6} C_{1} C_{3}^3 \left(5 \xi ^6+21 \xi
   ^2-39\right)+63 C_{3}^4 \left(7 \xi ^2-13\right)\right].
\end{eqnarray*}
With this corrected frequency at hand, one can just apply the same
method of resolution explained above to compute $\sigma_1$.


\section{Conclusions}\label{conclusions}

In this article, we have computed the power spectra of scalar and
tensor perturbations in a de Sitter universe using excited initial
states that arise from the invariants of a time-dependent harmonic
oscillator in the context of a full quantization of a perturbed FLRW
spacetime based on canonical quantum gravity with the Wheeler--DeWitt
equation. The quantization not only of the perturbations, but also of
the background spacetime, which leads to a Schr\"odinger equation with
quantum-gravitational correction terms, modifies the power spectra
with a specific $1/k^3$ correction. In the limit of the ground state,
the former results of Ref.~\cite{Brizuela2016a} for the corrected power spectra are recovered.

Given that such excited states also lead to higher-order correlations
contributing to the bispectrum of the perturbations, a further project
would be to study non-gaussianities arising from quantum-gravity
corrections in this context. 
Apart from that, the study of more general states, in particular,
coherent states would also be of interest. By considering a wide class of
initial states, one can obtain different results for the power
spectra, which would lead to different observational features. Whether
one can encounter a situation for which the tiny quantum-gravitational
correction terms are really measurable, is a big and important open issue.


\section*{Acknowledgments}

D.\,B.~acknowledges financial support from project FIS2017-85076-P
(MINECO/AEI/FEDER, UE) and Basque Government Grant No.~IT956-16. The
research of M.\,K.~was financed by the European Research Council Grant
No.~ERC-2013-CoG 616732 HoloQosmos.  He also thanks the \emph{Department of Theoretical Physics and History of Science} of the University of the Basque Country (UPV/EHU) for kind hospitality, while this work was finalized.
This paper is based upon work from COST action CA15117 (CANTATA),
supported by COST (European Cooperation in Science and Technology).

\appendix
\section{Computation of $\beta_{N_k}$}  
\subsection{$N_k$ even}

For the case $N_k$ even, the equation for $\widetilde y_1$ is
\begin{equation}
\frac{{\rm d}^2\widetilde y_1}{{\rm d}\xi^2}+\left(1-\frac{2}{\xi^2} \right)\widetilde y_1=
\frac{e^{-i \xi } \xi ^3 \left(11-\xi ^2\right)}{2 (\xi -i)^2 (\xi +i)^3}.
\end{equation}
and the initial conditions are
\begin{equation}
\widetilde\sigma^2_1\rightarrow-\frac{1}{4},\qquad (\widetilde\sigma^2_1)'\rightarrow 0.
\end{equation}
By solving this system one gets the following form for the mode,
\begin{eqnarray}
 \widetilde y_1&=&
\frac{\left(25-3 e^4\right) \pi}{8 e^2 \xi }  (\xi  \sin \xi+\cos\xi)\\\nonumber
&-&\frac{e^{-i\xi}}{8 \xi 
   (\xi +i)} \left[2 i \xi ^3+\xi ^2-14 i \left(\xi ^2+1\right) \arctan\xi+8 i \xi +1\right]\\\nonumber
&-&\frac{e^{i\xi}}{8 e^2 \xi }(\xi +i) \left[3 e^4 \text{Ci}(2 i-2 \xi )-7 \text{Ci}(2 \xi +2 i)\right.\\\nonumber
&+&\left.16 \text{Ei}(2-2 i \xi )+3 i e^4 \text{Si}(2 i-2 \xi )+7 i \text{Si}(2 \xi +2 i)\right],
\end{eqnarray}
with $\text{Si}$ and $\text{Ci}$ are the sine and cosine integral functions,
respectively, and $\text{Ei}$ the exponential integral function.
With this result one computes the correction number:
\begin{equation}
\beta_{N_k}=\frac{1}{4 e^2}\left[3 e^4 \text{Ei}(-2)+9 \text{Ei}(2)-e^2\right]\approx0.9876.
\end{equation}

\vskip 1mm
\subsection{$N_k$ odd}

For the case $N_k$ odd, the equation for $\widetilde y_1$ is
\begin{equation}
\frac{{\rm d}^2\widetilde y_1}{{\rm d}\xi^2}+\left(1-\frac{2}{\xi^2} \right)\widetilde y_1=
 \frac{e^{-i \xi } \xi ^3 \left(13-7 \xi ^2\right)}{6 (\xi -i)^2 (\xi +i)^3}.
\end{equation}
and the initial conditions are
\begin{equation}
\widetilde\sigma^2_1\rightarrow-\frac{7}{12},\qquad (\widetilde\sigma^2_1)'\rightarrow 0.
\end{equation}
By solving this system one gets the following form for the mode,
\begin{eqnarray}
 \widetilde y_1&=&\frac{\left(63-5 e^4\right) \pi }{24 e^2 \xi }(\xi  \sin\xi+\cos\xi)
\\\nonumber
&-&\frac{e^{-i\xi}}{24 \xi  (\xi +i)}\bigl[14 i \xi ^3+7 \xi ^2-34 i \left(\xi ^2+1\right)
\arctan\xi \\\nonumber
&~&\qquad\qquad\qquad+24 i \xi +7\bigr]\\\nonumber
&-&\frac{e^{i\xi}}{24 e^2 \xi }(\xi +i)\left[5 e^4 \text{Ci}(2 i-2 \xi )-33 \text{Ci}(2 \xi +2 i)\right.
\\\nonumber
&+&\left. 48 \text{Ei}(2-2 i \xi )
+5 i e^4\text{Si}(2 i-2 \xi )+33 i \text{Si}(2 \xi +2i) \right].
\end{eqnarray}
Replacing this functional form in the definition of $\widetilde\sigma_1$
and computing the limit, one gets the correction number:
\begin{equation}
\beta_{N_k}=\frac{1}{12 e^2} \left[5 e^4 \text{Ei}(-2)+15 \text{Ei}(2)-7 e^2\right]\approx0.104.
\end{equation}


\end{document}